\documentclass[a4paper]{article}

\usepackage[margin=1in]{geometry} 

\usepackage{amsmath}
\usepackage{amsthm}
\usepackage{amsfonts,amssymb}
\usepackage{bm}
\usepackage{bbm}
\usepackage{subcaption}
\usepackage{ulem}
\usepackage[utf8]{inputenc}
\usepackage[colorlinks,citecolor=blue,urlcolor=blue]{hyperref}

\usepackage[authoryear]{natbib}
\theoremstyle{plain}
\newtheorem{theorem}{Theorem}

\newtheorem{assumption}{Assumption}

\theoremstyle{definition}

\usepackage{graphicx,xcolor}

\usepackage{ algpseudocode} 
\usepackage{mathrsfs} 
\usepackage{algorithm2e}
\RestyleAlgo{ruled}

\usepackage{tikz}
\usetikzlibrary{arrows.meta, positioning}

\usepackage{lipsum}
\usepackage{multicol}
\RequirePackage{multirow}
\RequirePackage{booktabs}

\SetKwInput{KwInput}{Input}
\SetKwInput{KwOutput}{Output}
\SetKw{KwSet}{Set}
\SetKw{Break}{break}
\SetKwInput{KwModuleOne}{Module 1}
\SetKwInput{KwModuleTwo}{Module 2}
\SetKwInput{KwModuleThree}{Module 3}
\SetKwInput{KwModuleFour}{Module 4}
\SetKwComment{Comment}{/* }{ */}

\newcommand{\intervention}{\mathrm{int}}

\newcommand{\frake}{\mathfrak{e}}

\newcommand{\rmd}{\mathrm{d}}

\newcommand{\bbE}{\mathbb{E}}

\newcommand{\bbP}{\mathbb{P}}

\newcommand{\bbone}{\mathbbm{1}}

\newcommand{\calB}{\mathcal{B}}

\newcommand{\calM}{\mathcal{M}}
\newcommand{\calN}{\mathcal{N}}
\newcommand{\calO}{\mathcal{O}}

\newcommand{\calT}{\mathcal{T}}

\newcommand{\scrT}{\mathscr{T}}

\newcommand\independent{\protect\mathpalette{\protect\independenT}{\perp}}
\def\independenT#1#2{\mathrel{\rlap{$#1#2$}\mkern2mu{#1#2}}}

\title{\bf A flexible Bayesian g-formula for causal survival analyses with time-dependent confounding}
\author{Xinyuan Chen$^1$\thanks{xchen@math.msstate.edu} \and Liangyuan Hu$^2$\thanks{lh707@sph.rutgers.edu} \and Fan Li$^{3,4}$\thanks{fan.f.li@yale.edu}}

\date{
	$^1$Department of Mathematics and Statistics, Mississippi State University \\%
	$^2$Department of Biostatistics and Epidemiology, Rutgers School of Public Health \\%
	$^3$Department of Biostatistics, Yale University School of Public Health \\%
	$^4$Center for Methods in Implementation and Prevention Science, Yale University School of Public Health \\[2ex]%
}

\begin{document}
\maketitle

\begin{abstract}
	In longitudinal observational studies with time-to-event outcomes, a common objective in causal analysis is to estimate the causal survival curve under hypothetical intervention scenarios. The g-formula is a useful tool for this analysis. To enhance the traditional parametric g-formula, we developed an alternative g-formula estimator, which incorporates the Bayesian Additive Regression Trees (BART) into the modeling of the time-evolving generative components, aiming to mitigate the bias due to model misspecification. We focus on binary time-varying treatments and introduce a general class of g-formulas for discrete survival data that can incorporate the longitudinal balancing scores. The minimum sufficient formulation of these longitudinal balancing scores is linked to the nature of treatment strategies, i.e., static or dynamic. For each type of treatment strategy, we provide posterior sampling algorithms. We conducted simulations to illustrate the empirical performance of the proposed method and demonstrate its practical utility using data from the Yale New Haven Health System's electronic health records.\\
	
\end{abstract}


\section{Introduction} \label{sec:intro}

In longitudinal observational studies with time-to-event outcomes, a common goal of causal analysis is to characterize the causal survival curve under a specific (hypothetical) treatment strategy in the study population \citep{Hernan2020}. The challenge of drawing causal inferences in such studies arises from the complexity of observed time-varying treatments. These treatments are often influenced by past treatments and time-varying confounders, and in turn, they impact future treatment decisions and the event outcome. Furthermore, the time-to-event outcome can be subject to censoring due to loss to follow-up, leading to unobserved future treatments and event outcomes \citep{Keil2014}. The generalized computation formula, referred to as the g-formula \citep{Robins1986,Robins1987}, is an effective technique for determining standardized outcome distributions by adjusting for the treatment-confounder feedback.

Although the original form of the g-formula is nonparametric, its parametric implementation, based on the maximum likelihood estimates of its component models, has been more commonly used to ensure feasible computation. For instance, by employing generalized linear models, the parametric g-formula estimator has been implemented to adjust for a relatively large number of confounders with either non-censored \citep{Westreich2012} or censored outcomes \citep{Keil2014}. However, the parametric g-formula estimator faces two potential challenges. First, for studies involving a moderate number of time-varying confounders, modeling the entire history of time-varying confounders can be a daunting task, especially when the time-varying confounders involve a mix of continuous and categorical variables \citep{Achy-Brou2010}. Second, the consistency of the parametric g-formula estimator requires the correct specification of all component models, whereas the need for modeling multi-dimensional confounders in each period is susceptible to model misspecification \citep{daniel2013methods}. For example, overly parsimonious models can lead to the g-null paradox \citep{robins2013estimation} and introduce bias in the final causal effect estimates; see \citet{McGrath2022} for further discussion.  To this end, methods that improve the parametric g-formula with attention to more flexible model specifications are of increasing interest.

In this work, we focus on longitudinal observational studies with binary time-varying treatments and a discrete-time survival outcome and propose a new g-formula designed to overcome the aforementioned limitations. \citet{Keil2014} and \citet{Young2011,young2014identification} provided examples that implement the parametric g-formula for survival outcomes, and \citet{Wen2021} compared the operating characteristics of several parametric g-formula estimators. Their standard non-iterative implementation requires modeling the entire list of time-varying confounders as functions of their history. An alternative is the iterative conditional expectation g-formula \citep{Wen2021}, which obviates the need to model the time-varying confounders but requires a nested regression implementation. For the iterative g-formula estimator, the outcome model in period $t$ depends on the imputed potential outcomes in period $t+1$, which depends on the model fit in period $t+1$. Each estimator has its own features. Typically, the non-iterative g-formula estimator requires fitting models for both the outcomes and confounders across periods in a generative fashion and is flexible enough to address any causal estimands. In contrast, the iterative g-formula estimator only requires fitting outcome regression models conditional on confounder and treatment histories backward in period and is typically restricted to address mean causal estimands (a generalization to quantile estimands discussed in \citet{cheng2022doubly}). To improve the practical applicability of the parametric non-iterative g-formula when dealing with multi-dimensional time-varying confounders in scenarios with non-censored outcomes, \citet{Achy-Brou2010} and \citet{Shinohara2013} introduced a modified version of the parametric g-formula. This adaptation utilizes the longitudinal propensity scores to simplify the process.

The longitudinal propensity score defines the conditional probability of receiving the treatment in each period, balances the time-varying confounders, and is arguably easier to model compared to the multi-dimensional confounders \citep{Zhou2019}. This leads to our first contribution, which is to expand the non-iterative survival g-formula in \citet{Wen2021} to accommodate longitudinal balancing scores, with an overall objective of providing a useful alternative to the classic survival parametric g-formula. We show that the finest longitudinal balancing scores are the original confounders, and therefore, our version of the g-formula encompasses the one in \citet{Wen2021} as a special case. Conversely, the coarsest longitudinal balancing score is the product of the longitudinal propensity score and the censoring score, a scalar easier to model in each period and offers convenience and feasibility in implementation.

Our second contribution is to integrate the Bayesian Additive Regression Trees (BART), a flexible machine learner, into the survival g-formula to improve upon the standard implementation of the survival parametric g-formula estimator in longitudinal observational studies. Specifically, we incorporate BART for component models in a g-formula to mitigate the bias arising from parametric model misspecification \citep{Chipman2010,Sparapani2016,Tan2019}. We consider this approach because causal estimators based on BART have been empirically demonstrated to outperform competitors in drawing causal inference about the average treatment effect \citep{Hill2011}, conditional average treatment effect \citep{Henderson2018,Dorie2019,Hu2021,Hahn2020,caron2022estimating}, as well as causal effects with an intermediate variable  \citep{Josefsson2021,Bargagli2022,linero2022mediation,Chen2024}. This BART approach is flexible as it also provides full posterior samples for all unknown parameters, with built-in uncertainty quantification \citep{woody2021model}. 

The organization of the rest of this article is as follows. Section \ref{sec:not-asp-gform} provides an overview of the notation, data structure, and assumptions required for point identification of the population-level causal effects and describes the existing g-formula for causal survival analyses. Section \ref{sec:new-gform-thm} proposes a more general survival g-formula based on longitudinal balancing scores and offers key theoretical results to allow for subsequent simplification. Section \ref{sec:est-method} introduces BART into the proposed survival g-formula and describes the algorithms for drawing posterior samples for the target estimand. To illustrate the use of the proposed method, Section \ref{sec:num} includes a simulation study and an analysis of the Yale New Haven Health System electronic health record data. Section \ref{sec:dis} offers brief concluding remarks and outlines future extensions. The example \texttt{R} code for implementing the proposed estimators is available at the GitHub repository \url{https://github.com/erxc/gform}.

\section{Notation, assumptions and g-formula for discrete survival data} \label{sec:not-asp-gform}

\subsection{Notation and setup}

We consider a discrete-time longitudinal observational or cohort study comprising $N$ individuals. An individual's baseline ($t=0$) is defined as the period when they first satisfy all eligibility criteria to receive treatments. For each individual with a defined baseline, let $t \in \scrT=\{0,\ldots,T\}$ denote the collection of periods encompassing both the baseline and subsequent follow-up periods. We use $t^*$ to represent a target measurement period in $\scrT$. We use $A_t\in\{0,1\}$ to denote the binary treatment assignment in period $t$ with 1 for treated and 0 for not treated, and $L_t$ as a vector of confounders measured in period $t$. As in most survival studies, where the primary event of interest is often subject to censoring, we let binary indicators $Y_t$ and $C_t$ denote the outcome event status (e.g., death or stroke) and censoring status (e.g., loss to follow-up or study termination) in period $t$ in the sense that $Y_t\in\{0,1\}$ is observed if $C_t=0$ (uncensored) and unobserved if $C_t=1$ (censored). In addition, the overline notation is adopted to denote the history of a random vector up to period $t$, e.g., the history of confounders until period $t$ is given by $\overline{L}_t=\{L_0,\dots,L_t\}$ and the history of treatment status is given by $\overline{A}_t=\{A_0,\dots,A_t\}$. We consider that $L_0$ includes all baseline covariates, which may also include baseline assessments of any time-varying confounders in $L_t$; in other words, $L_0$ may be of higher dimensions than $L_t$ ($t\geq 1$), but this does not affect the development of the general methodology.

\begin{figure}[htbp]
	\centering
	\begin{tikzpicture}[node distance=1.6cm, auto]
		\node (Lt) {$L_t$};
		\node[right=of Lt] (Ct1) {$C_{t+1}$};
		\node[right=of Ct1] (Lt1) {$L_{t+1}$};
		\node[right=of Lt1] (Ct2) {$C_{t+2}$};
		\node[above right=of Lt, xshift=-0.6cm] (At) {$A_t$};
		\node[right=of At] (Yt1) {$Y_{t+1}$};
		\node[right=of Yt1] (At1) {$A_{t+1}$};
		\node[right=of At1] (Yt2) {$Y_{t+2}$};
		\node[left=of Lt, xshift=1.4cm] (ldot1) {$\ldots\ldots$};
		\node[left=of At, xshift=0.8cm] (ldot2) {$\ldots\ldots$};
		\node[right=of Ct2, xshift=-1.4cm] (ldot3) {$\ldots\ldots$};
		\node[right=of Yt2, xshift=-1.4cm] (ldot4) {$\ldots\ldots$};
		
		\draw[->, black, line width=0.2mm] (Lt) -- (At);
		\draw[->, black, line width=0.2mm] (Lt) -- (Ct1);
		\draw[->,black, line width=0.2mm] (Lt) -- (Yt1);
		\draw[->, black, line width=0.2mm] (At) -- (Yt1);
		\draw[->, black, line width=0.2mm] (At) -- (Ct1);
		\draw[->, red, line width=0.2mm] (Ct1) -- (Yt1);
		\draw[->, red, line width=0.2mm] (Yt1) -- (At1);
		\draw[->, red, line width=0.2mm] (Yt1) -- (Lt1);
		\draw[->, red, line width=0.2mm] (Yt1) -- (Ct2);
		\draw[->, red, line width=0.2mm] (Ct1) -- (Lt1);
		\draw[->, red, line width=0.2mm] (Ct1) -- (At1);
		\draw[->,black, line width=0.2mm] (Lt1) -- (At1);
		\draw[->,black, line width=0.2mm] (At1) -- (Yt2);
		\draw[->,black, line width=0.2mm] (At1) -- (Ct2);
		\draw[->,black, line width=0.2mm] (Lt1) -- (Ct2);
		\draw[->,black, line width=0.2mm] (Lt1) -- (Yt2);
		\draw[->,red, line width=0.2mm] (Ct2) -- (Yt2);
		
		\draw[->, bend left=80,black, line width=0.2mm] (Lt) to (At1);
		\draw[->, bend left=90,black, line width=0.2mm] (Lt) to (Yt2);
		\draw[->, bend right=60,black, line width=0.2mm] (Lt) to (Lt1);
		\draw[->, bend right=80,black, line width=0.2mm] (Lt) to (Ct2);
		\draw[->, bend left=30,black, line width=0.2mm] (At) to (At1);
		\draw[->, bend left=40,black, line width=0.2mm] (At) to (Yt2);
		\draw[->, bend right=85,black, line width=0.2mm] (At) to (Lt1);
		\draw[->, bend right=95,black, line width=0.2mm] (At) to (Ct2);
		\draw[->, bend left=30,red, line width=0.2mm] (Yt1) to (Yt2);
		\draw[->, bend right=30,red, line width=0.2mm] (Ct1) to (Ct2);
		
	\end{tikzpicture}
	\caption{A snapshot of the directed acyclic graph depicting possible pathways from period $t$ to period $t+1$ in the presence of time-varying confounding by $L$ and selection bias due to $C$ and $Y$. Each red edge refers to a semi-deterministic path from $C$ or $Y$ to other variables; if $C=1$ (censored) or $Y=1$ (the event occurred), the subsequent variable is considered unmeasured.}
	\label{fig:DAG}
\end{figure}
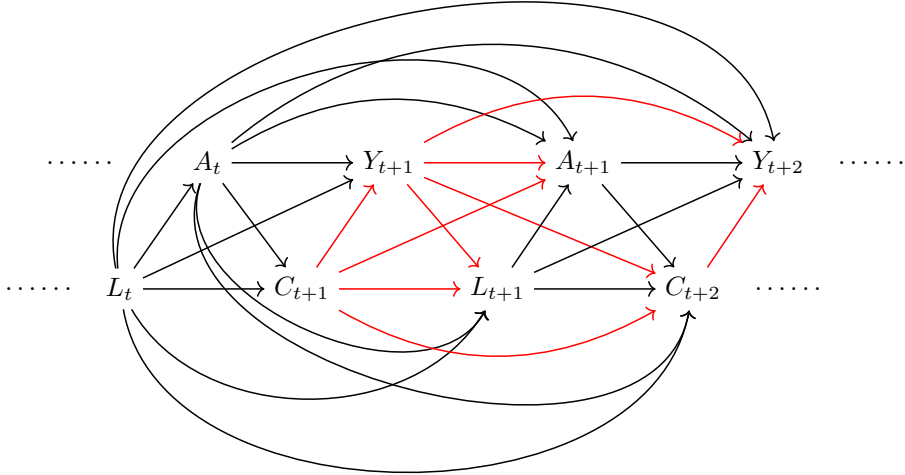

Similar to \citet{Young2011} and \citet{Wen2021}, we assume the quantities are observed in the order of $L_t\rightarrow A_t\rightarrow C_{t+1}\rightarrow Y_{t+1}$. Specifically, in period $t$, the confounder $L_t$ is observed with the assumption that it is affected by values of the past confounders $\overline{L}_{t-1}$, past treatments $\overline{A}_{t-1}$, event status $Y_t$, and censoring status $C_t$. The treatment assignment $A_t$ is then determined according to the treatment strategy, possibly in concert with $\overline{L}_{t-1}$, $\overline{A}_{t-1}$, and $Y_t$. The censoring status $C_{t+1}$ and outcome $Y_{t+1}$ in period $t+1$ are then sequentially observed based on values of $\overline{L}_t$, $\overline{A}_t$, $Y_t$, and $C_t$. This article focuses on the time to the first event, a common scenario in survival g-formula applications. We do not explore recurrent events and leave that extension for future work. When the primary event is death, it is natural that the follow-up will end for that individual after the event first occurred. If the primary event is non-terminal (e.g., stroke), we consider that the follow-up will terminate for that individual if the event occurred due to the interest in the time to the first event. Figure \ref{fig:DAG} provides a snapshot of the fully directed acyclic graph visualizing the relationships among variables in our longitudinal setup. To summarize, the follow-up terminates for an individual if either of the following scenarios occurs: (I) the primary event occurs before censoring: $Y_t=1$ and $C_t=0$ for some $t \in \scrT$, with the period $t$ recorded as the (discrete) event time; (II) the censoring occurs before the event: $Y_t$ is unobserved and $C_t=1$ for some $t \in \scrT$; (III) neither the primary event nor censoring occurs but the study period or maximum follow-up time is reached (administrative censoring): $Y_t=0$ and $C_t=0$ for all $t \in \scrT$. By definition, $Y_0=C_0=0$ and we define $\overline{L}_{-1}=\overline{A}_{-1}=\emptyset$ for notational convenience.

\subsection{Treatment strategies and causal estimands}\label{sec:estimand}

A \textit{treatment strategy} is a rule that assigns the treatment to individuals in each period $t \in \scrT$. Let $H_t\subseteq L_t$ denote a subset of confounders measured in period $t$. A treatment strategy can be described by its intervention distribution, $f^\intervention(a_t|\overline{h}_t,\overline{a}_{t-1},C_t=Y_t=0)$, assuming the potential dependence upon $(\overline{H}_t,\overline{A}_{t-1})=(\overline{h}_t,\overline{a}_{t-1})$. The subset of confounders, $H_t$, is often called the tailoring variables. Based on the structure of the intervention distribution, a strategy is labeled \emph{static} if its corresponding intervention distribution is independent of $\overline{H}_t$ and $\overline{A}_{t-1}$, and \emph{dynamic} otherwise \citep{Young2011,young2014identification,Wen2021}. This implies that, for a static strategy, the history of tailoring variables, $\overline{H}_t=\emptyset$, and for a dynamic strategy, $\emptyset\neq\overline{H}_t\subseteq\overline{L}_t$. In practice, for a dynamic strategy, while the dimension of time-varying confounders, $L_t$, can be large, tailoring variables that are used to inform treatment decisions of interest, $H_t$, may often include only a few biomarkers \citep{Hu2019}. 

Alternatively, a treatment strategy is \emph{deterministic} if $f^\intervention(a_t|\overline h_t,\overline a_{t-1},C_t=Y_t=0)$ is set to either 0 or 1 for all $t\in\scrT$, and \emph{random} otherwise. Naturally, deterministic and random strategies could be static or dynamic, and there are $2\times 2=4$ possible types of treatment strategies. In clinical studies, deterministic strategies are more frequently implemented compared to random strategies because of their simplicity in administrating treatments and interpreting treatment effects---two examples of deterministic strategies are the ``always treat'' strategy and the ``never treat'' strategy, corresponding to settings $a_t=1$ and $a_t=0$ $\forall~t\in\scrT$, respectively. \citet{Young2011} provided detailed examples of random treatment strategies, where $f^\intervention(a_t|\overline{h}_t,\overline{a}_{t-1},C_t=Y_t=0)$ may be considered to be the observed data treatment density for a given history with certain treatment monotonicity restrictions. Our discussions mainly center on deterministic treatment strategies due to their simplicity and ease of interpretation. Additionally, we will provide results for random treatment strategies as an expanded analysis. 

In this article, our interest lies in estimating the counterfactual risk of experiencing a key event (such as death) by period $t \in \scrT$ under pre-specified treatment strategies, whether static or dynamic, deterministic or random. We operate under the implicit assumption that the treatment strategies incorporate a hypothetical element designed to eliminate censoring during follow-up. To define the causal estimand, we adopt the notation in \citet{Wen2021} and first focus on deterministic treatment strategies. Specifically, we define $g\in\mathcal G$ as a deterministic treatment strategy, with $a_t^g=g_t(\overline{h}_t,\overline{a}^{g}_{t-1})$ being the assigned treatment in period $t$ under $g$. For any $g$, we define $Y_t^g$ and $L_t^g$ as the counterfactual outcome and confounder vector had an individual followed the strategy $g$; hence, the quantity $\bbE(Y_t^g)$ is defined as the counterfactual risk of experiencing the event of interest by period $t$ under $g$. Under a random strategy $f^\intervention(a_t|\overline{h}_t,\overline{a}_{t-1},C_t=Y_t=0)$, we use $\bbE^\intervention(Y_t^g)$ to represent the associated counterfactual risk estimand. As explained in \citet{Robins1986}, \citet{Young2011,young2014identification} and \citet{Wen2021}, $\bbE^\intervention(Y_t^g)=\sum_{g\in\mathcal{G}}\omega(g)\bbE(Y_t^g)$ is a weighted average of the estimand associated with a set of deterministic treatment strategies with weight $\omega(g)$ determined by the intervention distribution. To unify both cases, we use $\bbE^\intervention(Y_t^g)$ to generically represent the counterfactual risk estimand under a given intervention distribution, and $\bbE^\intervention(Y_t^g)=\bbE(Y_t^g)$ under a deterministic treatment strategy. For a given choice of $g$, it is then of interest to characterize the counterfactual discrete-time cumulative incidence function or the survival function---$\{\bbE^\intervention(Y_t^g), t\in\scrT\}$ or $\{\bbE^\intervention(1-Y_t^g), t\in\scrT\}$, and then compare these values under different choices of $g$ to study comparative effectiveness. 

Finally, we have primarily defined the super-population counterfactual risk estimand within a frequentist framework. Section \ref{sec:review} reviews the classical g-formula, which serves to identify this estimand. While frequentist implementations of the g-formula could, in principle, target the super-population estimand, a Bayesian implementation typically approximates it by targeting a mixed counterfactual risk estimand, as discussed in Section \ref{subsec:bayes-gform} (see also \citet{Li2022} for a related discussion in the context of a point treatment). In Section \ref{sec:sim}, we evaluate both frequentist and Bayesian g-formula methods under a frequentist framework, assuming the existence of a super-population estimand to ensure a common inferential target.

\subsection{Assumptions and the g-formula---a brief review}\label{sec:review}

We adopt the Neyman-Rubin \citep{Rubin1974} potential outcomes framework, which was generalized by \citet{Robins1986} to the longitudinal setting with time-varying treatments and confounding. Under this framework, to identify causal effects of time-varying treatment strategies, the following assumptions for each possible deterministic strategy $g\in\mathcal{G}$ of consideration are required: 
\begin{assumption}[\emph{Consistency}] \label{thm:asp-sutva}
		The outcomes and confounders of individual $i$ under treatment $\overline{a}_T=\{a_1,\ldots,a_T\}$ are independent of the assignment individual $j$ if $i\neq j$;
		furthermore, $\overline{Y}_{t+1}=\overline{Y}_{t+1}^g$ and $\overline{L}_t=\overline{L}_t^g$ if $\overline{A}_t=\overline{A}_t^g,\forall t\in\scrT$.
	\end{assumption}
	\begin{assumption}[\emph{Sequential ignorability}] \label{thm:asp-seq-ign}
		$\{Y_{t+1}^g,\dots,Y_T^g\}\independent \{A_t,C_{t+1}\}|\overline{L}_t=\overline{l}_t,\overline{A}_{t-1}=\overline{a}_{t-1}^g,\allowbreak C_t=Y_t=0,\forall~ t\in\scrT$.
	\end{assumption}
	\begin{assumption}[\emph{Positivity}] \label{thm:asp-pos}
		Let $f(C_{t+1}=0|\overline{l}_t,\overline{a}_t^g,C_t=Y_t=0)$ denote the conditional probability of no censoring in period $t+1$, and then the joint density of treatment status and no censoring, $f(a_t^g,C_{t+1}=0|\overline{l}_t,\overline{a}_{t-1}^g,C_t=Y_t=0)=f(a_t^g|\overline{l}_t,\overline{a}_{t-1}^g,C_t=Y_t=0)\times f(C_{t+1}=0|\overline{l}_t,\overline{a}_t^g,C_t=Y_t=0)>0$, whenever $f(\overline{L}_t=\overline{l}_t,\overline{A}_t=\overline{a}_t^g,C_t=Y_t=0)>0,\forall~ t\in\scrT$.
\end{assumption}

Assumption \ref{thm:asp-sutva} is also referred to as the Stable Unit Treatment Value Assumption (SUTVA), which excludes interference among individuals. It further assumes each treatment strategy involves well-defined intervention components without ambiguity, or otherwise variations in versions of the intervention components do not affect the counterfactual confounders and outcomes \citep{vanderweele2009concerning}. These conditions then allow us to connect the observed outcome and confounders with their counterfactual counterparts, depending on the realized treatment strategy. Assumption \ref{thm:asp-seq-ign}, also often referred to as conditional exchangeability \citep{Hernan2020}, states that the treatment assignment and censoring status of each individual in period $t$ are exogenous given the treatment and confounder histories, among those who remain in the study until period $t$ (neither the primary event nor censoring has occurred). Here, $\overline{L}_t$ could depend on the treatment strategy $g$, but we suppress this dependency for simplicity in stating the assumptions. Assumption \ref{thm:asp-pos} states that the conditional probability of censoring and the conditional probability of assignment is bounded away from 0 or 1 such that all counterfactual outcomes can be observed at each study period with a positive probability.

Under Assumptions \ref{thm:asp-sutva} - \ref{thm:asp-pos}, \citet{Robins1986} established the g-formula, which enables one to express $\bbE^\intervention(Y_{t^*}^g)$ as an expectation weighted by both the joint density of confounders and draws from the hypothetical intervention distribution \citep{Young2011,Wen2021}:
\begin{align} \label{eq:g-def-gen}
	\bbE^\intervention(Y_{t^*}^g)&=\int_{\overline{a}_{t^*-1}^g}\int_{\overline{l}_{t^*-1}}\sum_{t=1}^{t^*}\bbP(Y_t=1|\overline{L}_{t-1}=\overline{l}_{t-1},\overline{A}_{t-1}=\overline{a}_{t-1}^g,C_t=Y_{t-1}=0) \displaybreak[0]\nonumber \\
	&\quad\times \prod_{s=0}^{t-1}\bbP(Y_s=0|\overline{L}_{s-1}=\overline{l}_{s-1},\overline{A}_{s-1}=\overline{a}_{s-1}^g,C_s=Y_{s-1}=0) \displaybreak[0]\\
	&\quad\times f(l_s|\overline{l}_{s-1},\overline{a}_{s-1}^g,C_s=Y_s=0)f^\intervention(a_s|\overline{h}_s,\overline{a}_{s-1}^g,C_s=Y_s=0)\rmd\overline{l}_{t^*-1}\rmd\overline{a}_{t^*-1}^g, \nonumber
\end{align}
where $\bbP(Y_t=1|\overline{L}_{t-1}=\overline{l}_{t-1},\overline{A}_{t-1}=\overline{a}_{t-1},C_t=Y_{t-1}=0)$ and $f(l_t|\overline{l}_{t-1},\overline{a}_{t-1},\allowbreak C_t=Y_t=0)$ are the observed discrete-time hazards of the outcome and the joint density of confounders in period $t$, respectively, conditional on the histories of treatment and confounders, among those who survive through period $t-1$, and remain uncensored through period $t$. 

For deterministic treatment strategies, the integral over $\overline{a}_{t^*-1}^g$ in \eqref{eq:g-def-gen} is equivalent to plugging in the realization of treatment assignments $\overline{a}_{t^*-1}^g$. Thus, for a deterministic treatment strategy $g$, \eqref{eq:g-def-gen} can be simplified as \citep{Keil2014,Wen2021}
\begin{align} \label{eq:g-def-deter}
	\bbE^\intervention(Y_{t^*}^g)=\bbE(Y_{t^*}^g)&=\int_{\overline{l}_{t^*-1}}\sum_{t=1}^{t^*}\bbP(Y_t=1|\overline{L}_{t-1}=\overline{l}_{t-1},\overline{A}_{t-1}=\overline{a}_{t-1}^g,C_t=Y_{t-1}=0) \nonumber \displaybreak[0] \\
	&\quad\times \prod_{s=0}^{t-1}\bbP(Y_s=0|\overline{L}_{s-1}=\overline{l}_{s-1},\overline{A}_{s-1}=\overline{a}_{s-1}^g,C_s=Y_{s-1}=0)  \displaybreak[0]\\
	&\quad\times f(l_s|\overline{l}_{s-1},\overline{a}_{s-1}^g,C_s=Y_s=0)\rmd\overline{l}_{t^*-1}, \nonumber
\end{align}

The key step in estimating the counterfactual risk via \eqref{eq:g-def-gen} or \eqref{eq:g-def-deter} is to obtain estimates of the probabilities $\bbP(Y_t=1|\overline{L}_{t-1}=\overline{l}_{t-1},\overline{A}_{t-1}=\overline{a}_{t-1},C_t=Y_{t-1}=0)$, $\bbP(Y_s=0|\overline{L}_{s-1}=\overline{l}_{s-1},\overline{A}_{s-1}=\overline{a}_{s-1},C_s=Y_{s-1}=0)$, and $f(l_s|\overline{l}_{s-1},\overline{a}_{s-1},C_s=Y_s=0)$, for all necessary $t$ and $s$ involved in the g-formula, and then compute the integral over $\overline{l}_{t^*-1}$. When there are only a few binary or discrete confounders and study periods, it is possible to implement the g-formula using \eqref{eq:g-def-gen} or \eqref{eq:g-def-deter} nonparametrically. However, as the number of study periods increases, operationalizing the g-formula will quickly become unwieldy without parametric assumptions, even with a moderate number of time-varying confounders. Before detailing our implementation of the survival g-formula using Bayesian nonparametric priors---a strategy that relaxes parametric modeling assumptions to enhance the robustness of this complex estimator---we first describe a broader class of survival g-formula with increased flexibility in the modeling components. 

\section{A general class of survival g-formulas} \label{sec:new-gform-thm}

We introduce a more general class of g-formulas to estimate $\bbE^\intervention(Y_{t^*}^g)$ and present its theoretical grounds. Before we proceed, we first weaken the sequential ignorability assumption to facilitate the derivation of a g-formula with dimension reduction property.	
\begin{assumption}[\emph{Weak sequential ignorability}] \label{thm:asp-weak-seq-ign}
		$\{Y_{t+1}^g,\dots,Y_T^g\}\independent \bbone(A_t=a_t^g,C_{t+1}=0)|\overline{L}_t=\overline{l}_t,\overline{A}_{t-1}=\overline{a}_{t-1}^g,\allowbreak C_t=Y_t=0,\forall~ a_t^g\in\{0,1\}, t\in\scrT$.
\end{assumption}

Compared to the usual sequential ignorability assumption (Assumption \ref{thm:asp-seq-ign}), Assumption \ref{thm:asp-weak-seq-ign} only requires independence between the set of potential outcomes with an indicator function $\bbone(A_t=a_t^g,C_{t+1}=0)$ and is theoretically weaker (although both are untestable). It is straightforward to show that the survival g-formula presented in \eqref{eq:g-def-gen} can also be derived in the same fashion under Assumptions \ref{thm:asp-sutva}, \ref{thm:asp-pos} and \ref{thm:asp-weak-seq-ign}. In fact, when $a_t^g\in\{0,1\}$ and there is no censoring such that $C_{t+1}=0$ for all $t$, Assumption \ref{thm:asp-weak-seq-ign} and Assumption \ref{thm:asp-seq-ign} coincide. However, while the usual sequential ignorability assumption is only preserved if we condition on a multi-dimensional function of baseline and time-varying confounders, the weak sequential ignorability assumption continues to hold if we condition on a scalar function of baseline and time-varying confounders (an implication from Theorems \ref{thm:thm-bs} and \ref{thm:thm-seq-ign} below). This is analogous to the difference between strong and weak unconfoundedness in the setting of a time-fixed multi-valued treatment, as discussed in \citet{yang2016propensity}.

The proposed g-formula is anchored on the concept of longitudinal balancing scores, an extension of the balancing score first discussed in \citet{Rosenbaum1983}. To proceed, we first define the longitudinal balancing score, $b_t$, as a function of the history of confounders up to period $t$, $\overline L_t$, such that for each $\overline a_t$ generated by $g\in\mathcal G$,
\begin{align} \label{eq:bs-def}
	\{L_0,\ldots,L_t\}\independent \bbone(A_t=a_t,C_{t+1}=0)|\overline{b}_t,\overline{A}_{t-1}=\overline{a}_{t-1},\allowbreak C_t=Y_t=0,\forall t\in\scrT, 
\end{align}
where $\overline{b}_t=\{b_0,\ldots,b_t\}$ is the history of longitudinal balancing scores. The definition in \eqref{eq:bs-def} is an adaptation of the definition of cross-sectional balancing scores \citep{Rosenbaum1983} to the longitudinal observational study context, and, evidently, the most trivial longitudinal balancing scores are the entire history of observed confounders, $\overline{L}_t$. A more efficient use of \eqref{eq:bs-def} is to consider lower-dimensional functions of $\overline{L}_t$ as longitudinal balancing scores. Successfully substituting $\overline{L}_t$ with $\overline{b}_t$---particularly when $\overline{b}_t$ is of lower dimensionality than $\overline{L}_t$---can reduce computational complexity in the g-formula. 

One quantity closely related to the longitudinal balancing score is the longitudinal propensity score. We define the longitudinal joint propensity and censoring score, referred to as the longitudinal joint score, in period $t$ as 
\begin{align} \label{eq:long-ps}
	e_t=e_t(a_t|\overline{L}_t,\overline{a}_{t-1})=\bbP(A_t=a_t,C_{t+1}=0|\overline{L}_t,\overline{A}_{t-1}=\overline{a}_{t-1},C_t=Y_t=0),
\end{align}
which describes the likelihood of the individual assigned to $A_t=a_t$ while not being censored in period $t+1$ given the histories. The longitudinal joint score is a product of the propensity score $\bbP(A_t=a_t|\overline{L}_t,\overline{A}_{t-1}=\overline{a}_{t-1},C_t=Y_t=0)$ and the censoring score $\bbP(C_{t+1}=0|\overline{L}_t,A_t=a_t,\overline{A}_{t-1}=\overline{a}_{t-1},C_t=Y_t=0)$, which refers to the conditional probability of the individual remaining uncensored and event-free until period $t+1$, given the histories. With this definition, the relationship between the longitudinal balancing score and this joint score is then summarized in the following result:
\begin{theorem} \label{thm:thm-bs}
	Let $b_t$ be a function of $\overline{L}_t$. Then, $b_t$ is a longitudinal balancing score, that is, $\forall t\in\scrT$ and for each $\overline{a}_t$ generated by $g$,
	\begin{align*}
		\{L_0,\dots,L_t\}\independent \bbone(A_t=a_t,C_{t+1}=0)|\overline{b}_t,\overline{A}_{t-1}=\overline{a}_{t-1},\allowbreak C_t=Y_t=0,
	\end{align*}
	if and only if $b_t$ is finer than the longitudinal joint score $e_t$ in the sense that $e_t=\varphi(b_t)$ for some function $\varphi(\cdot)$ and $\forall t\in\scrT$.
\end{theorem}

Theorem \ref{thm:thm-bs} implies that the longitudinal joint score, $e_t$, is the coarsest longitudinal balancing score when the function $\varphi(\cdot)$ is set to be the identity function. This result is integral to computationally improving the g-formula since the longitudinal joint score is a scalar in each period $t$ and serves as a dimension reduction device. In addition, other longitudinal balancing scores could also be specified---e.g., the collection of tailoring variables plus the longitudinal joint score, $\{\overline{H}_t, \overline{e}_t\}$, as well as the combination of all confounders and the longitudinal joint score, $\{\overline{L}_t, \overline{e}_t\}$. Connecting our results to the cross-sectional point treatment setting, Theorem \ref{thm:thm-bs} reduces to Theorem 2 in \citet{Rosenbaum1983} if there is only a single period ($T = 1$) in the absence of censoring. The proof of Theorem \ref{thm:thm-bs} is presented in Web Appendix A.1. Next, we further establish the following result:
\begin{theorem} \label{thm:thm-seq-ign}
	Under Assumptions \ref{thm:asp-sutva}, \ref{thm:asp-pos} and \ref{thm:asp-weak-seq-ign}, weak sequential ignorability continues to hold when the history of confounders is substituted with the history of longitudinal balancing scores, that is, $\forall t\in\scrT$ and $g\in\mathcal{G}$, 
	$$\{Y_{t+1}^g,\ldots,Y_T^g\}\independent\bbone(A_t=a_t^g,C_{t+1}=0)|\overline{b}_t,\overline{A}_{t-1}=\overline{a}_{t-1}^g,C_t=Y_t=0.$$
\end{theorem}

Theorem \ref{thm:thm-seq-ign} demonstrates that weak sequential ignorability remains valid conditional on the history of longitudinal balancing scores, provided its counterpart is valid when conditioned on the history of confounders. This implies that one could leverage the longitudinal balancing scores to remove confounding by treating them as the original time-varying confounders. Similar connections to cross-sectional balancing scores can be found for Theorem \ref{thm:thm-seq-ign}, as it reduces to Theorem 3 in \citet{Rosenbaum1983} if there is only a single period ($|\scrT|=1$) without censoring. Furthermore, in the absence of censoring such that $\bbP(C_{t+1}=0|\overline{L}_t,A_t=a_t,\overline{A}_{t-1}=\overline{a}_{t-1},C_t=Y_t=0)=1$ for all $t\in\scrT$, Theorem \ref{thm:thm-seq-ign} reduces to Result 1 in \citet{Achy-Brou2010}. The proof of Theorem \ref{thm:thm-seq-ign} is provided in Web Appendix A.2. Importantly, Theorem \ref{thm:thm-seq-ign} suggests the possibility of expanding the existing survival g-formula to a more general class of g-formulas that act upon longitudinal balancing scores.
\begin{theorem} \label{thm:thm-new-gform}
	Under Assumptions \ref{thm:asp-sutva}, \ref{thm:asp-pos} and \ref{thm:asp-weak-seq-ign}, the expected outcome by period $t^*\in\scrT$ if all units had been assigned according to a treatment strategy defined by the intervention distribution $f^\intervention(a_s|\overline{h}_s,\overline{a}_{s-1},\allowbreak C_s=Y_s=0)$ can be identified:
	\begin{align} \label{eq:new-gform}
		\bbE^\intervention(Y_{t^*}^g)&=\int_{\overline{a}_{t^*-1}}\int_{\overline{b}_{t^*-1}}\sum_{t=1}^{t^*}\bbP(Y_t=1|\overline{b}_{t-1},\overline{A}_{t-1}=\overline{a}_{t-1},C_t=Y_{t-1}=0) \displaybreak[0]\nonumber \\
		&\quad\times \prod_{s=0}^{t-1}\bbP(Y_s=0|\overline{b}_{s-1},\overline{A}_{s-1}=\overline{a}_{s-1},C_s=Y_{s-1}=0) \displaybreak[0]\\
		&\quad\times f(b_s|\overline{b}_{s-1},\overline{a}_{s-1},C_s=Y_s=0)f^\intervention(a_s|\overline{h}_s,\overline{a}_{s-1},C_s=Y_s=0)\rmd\overline{b}_{t^*-1}\rmd\overline{a}_{t^*-1}. \nonumber
	\end{align}
	where $\overline{b}_s$ is the set of longitudinal balancing scores that at least include the history of tailoring variables $\overline{h}_s$ required by the intervention distribution. 
\end{theorem}

The proof of Theorem \ref{thm:thm-new-gform} can be found in Web Appendix A.3. Theorem \ref{thm:thm-new-gform} is a direct implication of Theorem \ref{thm:thm-seq-ign}. Evidently, the new g-formula in \eqref{eq:new-gform} reduces to the existing g-formula in \eqref{eq:g-def-gen} when $\overline{b}_s=\overline{L}_s$; however, \eqref{eq:new-gform} is more general as there are different possible specifications of the longitudinal balancing scores. Under the deterministic strategy, \eqref{eq:new-gform} simplifies to
\begin{align} \label{eq:new-gform-deter}
	\bbE^\intervention(Y_{t^*}^g)=\bbE(Y_{t^*}^g) &=\int_{\overline{b}_{t^*-1}}\sum_{t=1}^{t^*}\bbP(Y_t=1|\overline{b}_{t-1},\overline{A}_{t-1}=\overline{a}_{t-1}^g,C_t=Y_{t-1}=0) \displaybreak[0]\nonumber \\
	&\quad\times \prod_{s=0}^{t-1}\bbP(Y_s=0|\overline{b}_{s-1},\overline{A}_{s-1}=\overline{a}_{s-1}^g,C_s=Y_{s-1}=0) \displaybreak[0]\\
	&\quad\times f(b_s|\overline{b}_{s-1},\overline{a}_{s-1}^g,C_s=Y_s=0)\rmd\overline{b}_{t^*-1}. \nonumber
\end{align}

Different specifications of the longitudinal balancing scores can potentially serve different purposes. For example, one can choose the longitudinal joint score or any bijective mapping of these scores as the longitudinal balancing score in \eqref{eq:new-gform}. This approach could be preferable in some cases because $e_s$ is a scalar summary of the higher-dimensional $L_s$ in each period $s$, and it is substantially easier to model and then integrate over the one-dimensional longitudinal joint score than multi-dimensional confounders. For a dynamic treatment strategy, one could set $\overline{b}_s=\{\overline{H}_s, \overline{e}_s\}$ and maintain a lower dimensionality relative to $\overline{L}_s$, since the tailoring variables typically constitute a smaller subset of $\overline{L}_s$. In addition, \eqref{eq:new-gform} suggests an approach to consider both the longitudinal joint score and the full confounder by setting $\overline{b}_s=\{\overline{L}_s, \overline{e}_s\}$---the augmented confounder approach. As pointed out in \citet{zigler2013model} and \citet{Li2022}, this specification implicitly conducts outcome regression within each propensity score strata and has been referred to as a ``Bayesian analog of a doubly robust estimator" (even though it does not strictly possess the frequentist double robustness asymptotic property). In the time-fixed treatment setting, this specification has been shown to improve the efficiency for estimating causal effects compared to adjusting for confounders alone \citep{Hu2021}. Finally, Theorem \ref{thm:thm-new-gform} reduces to the g-formula for uncensored longitudinal observational data with deterministic static strategy in \citet{Achy-Brou2010} as a special case when we set $\overline{b}_s$ to the history of longitudinal propensity scores.  

\section{Bayesian inference for causal effects of treatment strategies} \label{sec:est-method}

We first provide a description of the Bayesian g-formula estimation framework in Section \ref{subsec:bayes-gform}, and then give the BART specification of component models in Section \ref{subsec:bart-spec}. Posterior inference procedures for the BART survival g-formula estimator are summarized in Section \ref{subseq:post-inf}.

\subsection{The Bayesian implementation of g-formula} \label{subsec:bayes-gform}

For notational simplicity, we suppress individual subscripts and let $\calO=\{L,b,C,A,Y\}$ denote the collection of observables (the balancing score is an estimable quantity, and we include it in $\calO$), and, we use $p(\cdot)$ to denote a generic distribution. Under a given treatment strategy defined by the intervention distribution, the Bayesian approach infers $\bbE^\intervention(Y_{t^*}^g)$ through the following two steps: (I) Draw samples from $p^\intervention({Y}_{t^*}^g|\calO)$, the posterior predictive distribution (PPD) of ${Y}_{t^*}^g$ under $f^\intervention(a_t|\overline{h}_t,\overline{a}_{t-1},C_t=Y_t=0)$, which marginalizes over posterior distributions of the unknown parameters; and (II) Summarize draws from the PPD to obtain point and interval estimates.

Specifically, for a deterministic treatment strategy based on \eqref{eq:new-gform-deter}, two sets of generative models are involved when implementing the Bayesian approach: (I) the outcome models $p(Y_t|\overline{b}_{t-1},\overline{A}_{t-1}=\overline{a}_{t-1}^g,C_t=Y_{t-1}=0)$ for $t=1,\ldots,t^*$, which are also referred to as the discrete-time hazard function, and (II) longitudinal balancing score models $p(b_t|\overline{b}_{t-1},\overline{A}_{t-1}=\overline{a}_{t-1}^g,C_t=Y_t=0)$, for $t=1,\ldots,t^*-1$. Below, we denote these models by $p(Y_t|\overline{b}_{t-1},\overline{a}_{t-1}^g,\beta_t)$ and $p(b_t|\overline{b}_{t-1},\overline{a}_{t-1}^g,\eta_t)$, respectively, where the surviving and uncensored condition, $C_t=Y_{t-1}=0$, is omitted for brevity, and $(\beta_t,\eta_t)$ are model parameters that will be assumed \emph{a priori} independent. Let $\overline{\beta}_t=\{\beta_1,\ldots,\beta_t\}$ and $\overline{\eta}_t=\{\eta_1,\ldots,\eta_t\}$, and then the likelihood function for ${Y}_{t^*}^g$ under $g\in\mathcal{G}$ is
\begin{align*}
	p({Y}_{t^*}^g|\overline{\beta}_{t^*},\overline{\eta}_{t^*-1})=&\int_{\overline{b}_{t^*-1}}\sum_{t=1}^{t^*}p({Y}_t=1|\overline{b}_{t-1},\overline{a}_{t-1}^g,\beta_t)\displaybreak[0]\\
	&\quad\times\prod_{s=0}^{t-1}p({Y}_s=0|\overline{b}_{s-1},\overline{a}_{s-1}^g,\beta_s)p(b_s|\overline{b}_{s-1},\overline{a}_{s-1}^g,\eta_s)\rmd\overline{b}_{t^*-1}.
\end{align*}

Assuming parameters for different models are \emph{a priori} independent with priors, $p(\overline{\beta}_{t^*})$ and $p(\overline{\eta}_{t^*-1})$, the estimation of posteriors, $p(\overline{\beta}_{t^*}|\calO)$ and $p(\overline{\eta}_{t^*-1}|\calO)$, can be modularized so that the posterior distributions for these two sets of models are inferred through distinct and corresponding Bayesian procedures \citep{Keil2018}. Subsequently, the PPD under $g$ is obtained through
\begin{align} \label{eq:det-ppd}
	p({Y}_{t^*}^g|\calO)=\int_{\overline{\beta}_{t^*}}\int_{\overline{\eta}_{t^*-1}}p({Y}_{t^*}^g|\overline{\beta}_{t^*},\overline{\eta}_{t^*-1})p(\overline{\beta}_{t^*}|\calO)p(\overline{\eta}_{t^*-1}|\calO)\rmd\overline{\eta}_{t^*-1}\rmd\overline{\beta}_{t^*}.
\end{align}
Building on this idea, for a random treatment strategy pinned down by $f^\intervention(a_t|\overline{h}_t,\overline{a}_{t-1})$, the likelihood function is updated as
\begin{align*}
	p^\intervention({Y}_{t^*}^g|\overline{\beta}_{t^*},\overline{\eta}_{t^*-1})=&\int_{\overline{a}_{t^*-1}}\int_{\overline{b}_{t^*-1}}\sum_{t=1}^{t^*}p({Y}_t=1|\overline{b}_{t-1},\overline{a}_{t-1},\beta_t)\displaybreak[0]\\
	&\quad\times\prod_{s=0}^{t-1}p({Y}_s=0|\overline{b}_{s-1},\overline{a}_{s-1},\beta_s)p(b_s|\overline{b}_{s-1},\overline{a}_{s-1},\eta_s)\displaybreak[0]\\
	&\quad\times f^\intervention(a_s|\overline{h}_s,\overline{a}_{s-1})\rmd\overline{b}_{t^*-1}\rmd\overline{a}_{t^*-1},
\end{align*}
and the PPD is $p^\intervention({Y}_{t^*}^g|\calO)=\int_{\overline{\beta}_{t^*}}\int_{\overline{\eta}_{t^*-1}}p^\intervention({Y}_{t^*}^g|\overline{\beta}_{t^*},\overline{\eta}_{t^*-1})p(\overline{\beta}_{t^*}|\calO)p(\overline{\eta}_{t^*-1}|\calO)\rmd\overline{\eta}_{t^*-1}\rmd\overline{\beta}_{t^*}$. 

The subsequent step in implementing the proposed g-formula involves inferring the posterior distributions of the longitudinal joint score. We follow a similar methodology to that proposed in \citet{Hahn2020}, which includes first obtaining posterior distributions of longitudinal joint scores via a Bayesian procedure. Subsequently, we use the posterior means of these longitudinal joint scores as covariates in the outcome and longitudinal balancing score models. Specifically,  the longitudinal joint score in \eqref{eq:long-ps} is $e_t = \bbP(A_t=a_t|\overline{l}_t,\overline{a}_{t-1},C_t=Y_t=0)\times\bbP(C_{t+1}=0|\overline{l}_t,\overline{a}_t,C_t=Y_t=0)$. Omitting the surviving and uncensored condition, we assume models for the observed data, $p(A_t=1|\overline{l}_t,\overline{a}_{t-1},\xi_t)$ and $p(C_{t+1}=0|\overline{l}_t,\overline{a}_t,\zeta_t)$, with \emph{a priori} independence between $\overline{\xi}_t=\{\xi_1,\ldots,\xi_t\}$ and $\overline{\zeta}_t=\{\zeta_1,\ldots,\zeta_t\}$ (priors: $p(\overline{\xi}_{t^*-1})$ and $p(\overline{\zeta}_{t^*})$). We then obtain the estimate of $e_t$ as $\widehat{e}_t = \widehat{p}(A_t=a_t|\overline{l}_t,\overline{a}_{t-1},\xi_t)\times\widehat{p}(C_{t+1}=0|\overline{l}_t,\overline{a}_t,\zeta_t)$ for each individual based on their observed treatment path and censoring pattern. Here, $\widehat{p}(A_t=a_t|\overline{l}_t,\overline{a}_{t-1},\xi_t)$ and $\widehat{p}(C_{t+1}=0|\overline{l}_t,\overline{a}_t,\zeta_t)$ are the respective posterior mean probabilities, based on which $\widehat{e}_t$ will be incorporated as a time-varying covariate in the g-formula. 

This method, which involves conditioning on the inferred posterior distributions of longitudinal joint scores, is not subject to the feedback issue \citep{Zigler2014}. This is because the model parameters $\xi_t$ and $\zeta_t$ are utilized exclusively during inferring longitudinal joint scores. Additionally, this approach can be considered as an application of Zellner's $g$-prior \citep{Zellner1986}, where the prior covariance of a vector of regression coefficients is parameterized using a plug-in estimate of the predictor variables' covariance matrix. \citet{Hahn2020} justified this approach to include the estimated propensity scores in nonparametric Bayesian regression models to reduce regularization-induced confounding in estimating the heterogeneous treatment effects with a time-fixed treatment. 

Conceptually, While our Bayesian implementation of the survival g-formula requires explicit modeling time-varying confounders or balancing scores, the final estimator for $\bbE^\intervention(Y_{t^*}^g)$ still averages over the observed baseline confounder distribution. In this sense, our estimator targets a mixed counterfactual risk estimand, which is a natural and convenient approximation to the population counterfactual risk estimand $\bbE^\intervention(Y_{t^*}^g)$ (see \citet[\S3]{Li2022} for a discussion on mixed average treatment effect versus population average treatment effect). Finally, we notice that the purpose of modeling the longitudinal joint score is to facilitate dimension reduction. Since the longitudinal joint scores involve the treatment and censoring processes, our Bayesian implementation of the survival g-formula should be considered as a quasi-Bayesian approach because the observed data likelihood does not involve the treatment and censoring processes under weak sequential ignorability (see \citet{Li2022} for a discussion under a time-fixed treatment setting). That is, our implementation combines Bayesian inference for separate component models in the survival g-formula.

\subsection{Model specification via Bayesian Additive Regression Trees} \label{subsec:bart-spec}

To increase the flexibility and robustness of the estimation procedure, we propose to specify and infer the component models of the new g-formula using BART \citep{Chipman2010}. \citet{Hill2011} first introduced BART as a flexible Bayesian nonparametric procedure for estimating the average causal effect with a time-fixed treatment. Several subsequent simulation studies have shown that, compared to other machine learners, causal estimators derived from BART usually have the best performance metrics under a time-fixed treatment setting (see, for example, \citet{hu2020estimation} for binary outcomes and \citet{Hu2021} for survival outcomes, \citet{Dorie2019} for comparative results from causal inference data analysis challenge launched in the 2016 Atlantic Causal Inference Conference, and \citet{caron2022estimating} for a recent overview and comparison of methods to estimate conditional average causal effects). In addition, among several advancements in Bayesian nonparametric methods, e.g., the Dirichlet process prior \citep{Antonelli2019}, the Gaussian process prior \citep{zhu2023addressing}, and the Gamma process prior \citep{Oganisian2024}, BART has the additional advantage of computational efficiency with minimum tuning. For the concreteness of method development, we, therefore, describe procedures to integrate BART into addressing causal effects with a time-varying treatment effect but acknowledge that other Bayesian nonparametric priors can be similarly integrated into the proposed survival g-formula.

As an ensemble method, BART represents component models via sums of individual trees, with prior distributions imposed to regularize the fit by keeping the individual tree effects relatively small \citep{Chipman2010}. Specifically, let $\calT$ denote a binary tree consisting of interior node decision rules and terminal nodes, and let $\calM =\{\mu_1,\ldots,\mu_K\}$ denote parameters associated with each of the $K$ terminal nodes of $\calT$. The BART specification of each component model is developed upon a collection of $J$ binary trees $\{\calT_1,\ldots,\calT_J\}$ and their respectively associated set of terminal node parameters $\{\calM_1,\ldots,\calM_J\}$ for each tree, where $\calM_j =\{\mu_{j,1},\ldots,\mu_{j,K_j}\}$ is a collection of $K_j$ terminal node parameters. Each tree $\calT_j$ consists of a sequence of decision rules, through which any covariate vector can be assigned to one terminal node of $\calT_j$ according to the decision rules prescribed at each interior node. Particularly, the decision rules at the interior nodes of $\calT_j$ are of the form $\{X_q\leq c\}$ versus $\{X_q>c\}$, where $X_q$ denotes the $q$-th element of covariate vector $X$. A covariate vector $X$ corresponding to the $m$-th terminal node of $\calT_j$ is assigned the value $\mu_{j,m}$ and $u(X;\calT_j,\calM_j)$ is used to denote the function returning $\mu_{j,m} \in \calM_j$ whenever $X$ is assigned to the $m$th terminal node of $\calT_j$. A generic component model of the g-formula, $p(X;\theta)$, can thus be represented as a sum of individual trees
\begin{align*}
	p(X;\theta)=\sum_{j=1}^J u(X;\calT_j,\calM_j).
\end{align*}

Under BART, prior distributions on $(\calT_j,\calM_j)$ induce a prior on $u(X;\calT_j,\calM_j)$ and hence a prior on $p(X;\theta)$. To proceed, one needs to specify the following to complete the description of the prior on $\{(\calT_j,\calM_j), j=1,\ldots,J\}$: (I) the distribution on the choice of splitting variable at each internal node; (II) the distribution of the splitting value $c$ used at each internal node; (III) the probability that a node at a given node-depth $\delta$ splits, which is assumed to be equal to $\tau(1+\delta)^{-\alpha}$; and (IV) the distribution of the terminal node values $\mu_{j,m}$. Regarding (I) - (IV), we defer to defaults in \citet{Chipman2010}, where, for (I), the splitting variable is chosen uniformly from the set of available splitting variables at each interior node; for (II), a uniform prior on the discrete set of available splitting values is adopted; for (III), the depth-related hyperparameters are chosen as $\tau=0.95$ and $\alpha=2$. For (IV), the distribution of the terminal node values $\mu_{j,m}$ is assumed to be $\mu_{j,m}\sim \calN(0, (4w^2J)^{-1})$, where $w=2$ and $J=200$. We use the notation $p(\cdot;\theta)\sim \mathrm{BART}(\tau,\alpha,w,J)$ to denote the distribution on component model function $p(\cdot;\theta)$ induced by the prior distribution on $\calT_j$ and $\calM_j$ with parameter values $(\tau,\alpha,w)$ and $J$ total trees. 

We specify the binary outcome model, $p(Y_t|\overline{b}_{t-1},\overline{a}_{t-1},\beta_t)$, using logit BART \citep{sparapani2021nonparametric} such that
\begin{align} \label{eq:bart-oc-mod}
	p(Y_t=1|\overline{b}_{t-1},\overline{a}_{t-1},\beta_t) = \Psi\left\{\sum_{j=1}^{J_{Y,t}} u_{Y,t}\left(\overline{b}_{t-1},\overline{a}_{t-1};\calT_j^{Y,t},\calM_j^{Y,t}\right)\right\},
\end{align}
where $\Psi(\cdot)$ is the standard logistic cumulative distribution function, and $\beta_t$ now includes the associated tree parameters. Though \eqref{eq:bart-oc-mod} provides a potentially saturated description for specifying the outcome model using BART, its practical implementation may encounter challenges. These include computational complexity and numerical instability, which arise from fitting time-varying tree components with the accumulating history of longitudinal balancing scores. In other words, the dimension of $\overline{b}_t$ could grow fast as $t$ increases, making the proposed method difficult to implement since a new BART model needs to be fitted for every period $t$. 

\citet{Robins2000} discussed a similar issue and suggested that one could simplify the implementation by formulating more parsimonious models with lower-dimensional functions of the cumulative history and/or specifying the order of dependence. For example, one might choose to use the cumulative sum of the history of longitudinal balancing scores together with a dependence of order two, i.e., the design vector concerning the balancing scores becomes $\overline{b}_t^*=(\sum_{s=0}^{t-3}b_s',b_{t-1}',b_{t-2}')'$. Hence, we can parameterize model \eqref{eq:bart-oc-mod} as
\begin{align} \label{eq:bart-oc-mod-simp}
	p(Y_t=1|\overline{b}_{t-1}^*,\overline{a}_{t-1},\beta_t) = \Psi\left\{\sum_{j=1}^{J_Y} u_Y\left(\overline{b}_{t-1}^*,\overline{a}_{t-1};\calT_j^Y,\calM_j^Y\right)\right\},
\end{align}
where the same set of tree components is fitted and shared across all study periods. This approach allows for the pooling of all observations from all periods, thereby reducing the computational burden and improving the finite-sample stability. Similarly, to estimate the longitudinal joint scores, we specify 
\begin{equation}\label{eq:bart-trt-cen}
	\begin{aligned} 
		p(A_t=1|\overline{l}_t,\overline{a}_{t-1},\xi_t) &= \Psi\left\{\sum_{j=1}^{J_A} u_A\left(\overline{l}_t^*,\overline{a}_{t-1};\calT_j^A,\calM_j^A\right)\right\}, \\
		p(C_{t+1}=0|\overline{l}_t,\overline{a}_t,\zeta_t) &= \Psi\left\{\sum_{j=1}^{J_C} u_C\left(\overline{l}_t^*,\overline{a}_t;\calT_j^C,\calM_j^C\right)\right\},
	\end{aligned}   
\end{equation}
where $\overline{l}_t^*$ is defined in the same fashion as $\overline{b}_t^*$ to enable feasible computation based on parsimonious model specification, and $\xi_t$ and $\zeta_t$ include the respective tree parameters. Based on these models, $\widehat{e}_t$ is derived from the product of posterior mean probabilities of component models in \eqref{eq:bart-trt-cen}.

We then specify the model for longitudinal balancing scores as time-varying covariates. We first transform the longitudinal joint score into its logit, $\frake_t = \log\{\widehat{e}_t/(1-\widehat{e}_t)\}\in\mathbb{R}=[-\infty,\infty]$, which eliminates the numerical constraint of $\widehat{e}_t\in[0,1]$ and is monotone in $\widehat{e}_t$. The logit of the longitudinal joint score, $\frake_t$, is viewed as a continuous covariate over $\mathbb{R}$ and can subsequently be modeled together with other elements of $b_t$ (or on its own if $b_t$ includes $e_t$ as a sole component). Here, we consider the setting of a limited number of elements that are either binary or continuous with no constraints, and covariates of other types can be converted into binary or continuous when modeling. An iterative BART approach is adopted for the model of $b_t$. For clarity, we describe this approach via an example, where $b_t=(b_{t,1},b_{t,2})'$ with $b_{t,1}$ and $b_{t,2}$ being binary and continuous respectively. Specifically, we have
\begin{equation} \label{eq:bart-long-bs-mod}
	\begin{aligned}
		p(b_{t,1}=1|\overline{b}_{t-1},\overline{a}_{t-1},\eta_{t,1}) &= \Psi\left\{\sum_{j=1}^{J_{b,1}} u_{b,1}\left(\overline{b}_{t-1}^*,\overline{a}_{t-1};\calT_j^{b,1},\calM_j^{b,2}\right)\right\}, \\
		p(b_{t,2}|b_{t,1},\overline{b}_{t-1},\overline{a}_{t-1},\eta_{t,2}) &= \calN\left(\sum_{j=1}^{J_{b,2}} u_{b,2}\left(b_{t,1},\overline{b}_{t-1}^*,\overline{a}_{t-1};\calT_j^{b,2},\calM_j^{b,2}\right),\sigma_{b,2}^2\right),
	\end{aligned}   
\end{equation}
where the distribution of $(b_{t,1}|\overline{b}_{t-1},\overline{a}_{t-1})$ is specified via the logit BART, the distribution of $(b_{t,2}|b_{t,1},\overline{b}_{t-1},\overline{a}_{t-1})$ is modeled by a normal distribution with the mean function specified via BART, and $\eta_{t,1}$ and $\eta_{t,2}$ are the relevant tree parameters. This iterative BART approach simplifies the procedure of modeling the correlation amongst elements of $b_t$, and additional conditional distributions could be included if $b_t$ contains more elements. It is applicable in practice because the longitudinal balancing score is usually lower-dimensional and, at the minimum, only contains the longitudinal joint score and the tailoring covariates. The iterative order amongst binary and continuous elements is not fixed and can be specified by the user.

\subsection{Posterior inference} \label{subseq:post-inf}

Based on the aforementioned BART specifications, the posterior inference for the causal effect under treatment strategy $g$ is achieved by sampling from the PPD of the potential outcomes, as described in \eqref{eq:det-ppd}. This process, in general, consists of four main modules, and we summarize each module below.
\begin{enumerate}
	\item Specify priors for BART propensity score and censoring score model parameters in \eqref{eq:bart-trt-cen}, $(\calT^A,\calM^A)$ and $(\calT^C,\calM^C)$; obtain the estimated longitudinal joint score. The logit of the longitudinal joint score will be included as an element of the longitudinal balancing score in subsequent computations.
	\item Specify priors for BART model parameters in \eqref{eq:bart-oc-mod-simp} and \eqref{eq:bart-long-bs-mod}, $(\calT^Y,\calM^Y)$ and $(\calT^b,\calM^b,\sigma_b^2)$, as required by each component models in the survival g-formula; obtain posterior samples of $(\calT^Y,\calM^Y)$ and $(\calT^b,\calM^b,\sigma_b^2)$ via Markov Chain Monte Carlo (MCMC) procedures described in \citet{Chipman1998} and \citet{sparapani2021nonparametric}. 
	\item Generate samples of ${Y}_{t^*}^g$ from its PPD in \eqref{eq:det-ppd} via Monte Carlo integration because the PPD typically is analytically intractable due to the compounding complexity from component models. Posterior samples of $(\calT^Y,\calM^Y)$ and $(\calT^b,\calM^b,\sigma_b^2)$ are used as model parameters in \eqref{eq:bart-oc-mod-simp} and \eqref{eq:bart-long-bs-mod} to generate potential outcomes and longitudinal balancing scores under the treatment strategy of interest. For a deterministic treatment strategy, setting the treatment values for each period is straightforward. For a random treatment strategy, one needs to draw from $f^\intervention$.
	\item Obtain posterior samples of the treatment effect by summarizing draws from the PPD.
\end{enumerate}

An outline of the pseudo-code describing the BART g-formula under a deterministic treatment strategy $g$ is summarized in Algorithm \ref{alg:gform-det}. Specifically, for computational feasibility, the numerical integration over the posterior distribution of model parameters in \eqref{eq:det-ppd} is simplified to using the value obtained from a single draw instead of the average over multiple draws. The posterior mean of the treatment effect can be obtained from samples in Module 4. Web Appendix B describes the extension to a random treatment strategy, where a numerical grid composed of samples of longitudinal balancing scores and treatment assignments is generated to incorporate the additional layer of integration over the history of treatment assignments. 

\begin{algorithm}[htbp]
	
	\caption{A BART approach for implementing the survival g-formula under a deterministic treatment strategy $g\in\mathcal G$.}\label{alg:gform-det}
	\KwInput{$\{l_t, a_t^g\}$ for $t=0,\ldots,T-1$, and $\{C_t, Y_t\}$ for $t=1,\ldots,T$.}
	\KwOutput{Posterior samples of $\bbE(Y_{t^*}^g)$.}
	\KwModuleOne
	
	Specify independent priors for $(\calT^A,\calM^A)$ and $(\calT^C,\calM^C)$ and fit logit BART models in \eqref{eq:bart-trt-cen}. Obtain posterior mean probabilities $\widehat{p}(A_t=1|\overline{l}_t,\overline{a}_{t-1},\xi_t)$ and $\widehat{p}(C_{t+1}=0|\overline{l}_t,\overline{a}_t,\zeta_t)$, and compute estimated longitudinal joint propensity and censoring score, $\widehat{e}_t=\widehat{p}(A_t=1|\overline{l}_t,\overline{a}_{t-1},\xi_t)\times\allowbreak\widehat{p}(C_{t+1}=0|\overline{l}_t,\overline{a}_t,\zeta_t)$, with corresponding logits, $\frake_t = \log\{\widehat{e}_t/(1-\widehat{e}_t)\}$.
	
	\KwModuleTwo
	
	Specify independent priors for $(\calT^Y,\calM^Y)$ and $(\calT^b,\calM^b,\sigma_b^2)$ and fit logit and continuous BART models in \eqref{eq:bart-oc-mod-simp} and \eqref{eq:bart-long-bs-mod} with $\frake_t$ included as an element of $b_t$. Obtain $R$ samples of BART parameters, i.e., $\left\{(\calT^Y,\calM^Y)^{(r)},(\calT^b,\calM^b,\sigma_b^2)^{(r)}\right\}_{r=1}^R$.
	
	\KwModuleThree
	
	\For{$r=1,\ldots, R$}{
		1. Generate a bootstrap sample of size $K$, $\left\{b_{0,k}^{(r)}\right\}_{k=1}^K$, from the empirical distribution of $b_0$, $\widehat{p}(b_0)$. 
		
		2. Assign treatment, $a_{0,k}^{g,(r)}$, to each of the $K$ samples according to the deterministic treatment strategy $g$. 
		
		\For{$k=1,\ldots,K$}{
			Set $t=1$.
			
			\While{$t\leq T$}{
				1. Generate ${Y}_{t,k}^{(r)}$ from \eqref{eq:bart-oc-mod-simp} with $(\calT^Y,\calM^Y)^{(r)}$ as model parameters, i.e., $p\left({Y}_{t,k}^{(r)}=1|\overline{b}_{t-1,k}^{(r)},\overline{a}_{t-1,k}^{g,(r)},\widehat{\beta}_t^{(r)}\right) = \Psi\left\{\sum_{j=1}^{J_Y} u_Y\left(\overline{b}_{t-1,k}^{*,(r)},\overline{a}_{t-1,k}^{g,(r)};\calT_j^{Y,(r)},\calM_j^{Y,(r)}\right)\right\}$.
				
				\eIf{${Y}_{t,k}^{(r)}=0$}{
					2. Generate $b_{t,k}^{(r)}$ from \eqref{eq:bart-long-bs-mod} with $(\calT^b,\calM^b)^{(r)}$ as model parameters.
					
					3. Assign treatment, $a_{t,k}^{g,(r)}$, according to the deterministic treatment strategy $g$. 
					
					4. Set $t=t+1$.
				}{
					\Break
				}
			}
		}
	}
	
	\KwModuleFour
	
	\For{$r=1,\ldots,R$}{
		Compute $\widehat\bbE({Y}^g_{t^*}){}^{(r)} = K^{-1}\sum_{k=1}^K\sum_{t=1}^{t^*}{Y}_{t,k}^{(r)}$ and obtain its posterior.
	}
	
\end{algorithm}

\section{Numerical examples} \label{sec:num}

\subsection{Simulation studies with synthetic data} \label{sec:sim}

We use simulations to evaluate the finite-sample performance of the proposed BART estimators with comparisons to other methods. We create a setting where $N=1,000$ individuals were included in a longitudinal observational study across $T=5$ periods. Three confounders are simulated, $L_t=(L_{t,1},L_{t,2},L_{t,3})'$, where $L_{t,1}$ is a binary variable, and $L_{t,2}$ and $L_{t,3}$ are two continuous variables. We assume the following generative models for the confounders, that is, for $t=1,\ldots,T-1$,
\begin{align*}
	L_{t,1}&\sim \calB\big(\Psi(-2A_{t-1}+0.2L_{t-1,1})\big)\big\rvert L_{t-1,1}, \displaybreak[0]\\
	L_{t,2}&\sim \calN\left(-2A_{t-1}+0.2L_{t-1,1}+L_{t-1,2}L_{t-1,3}+\sin(L_{t-1,2}),0.1^2\right),\displaybreak[0] \\
	L_{t,3}&\sim \calN\left(-2A_{t-1}+0.2L_{t-1,1}+L_{t-1,2}L_{t-1,3}+\sin(L_{t-1,3}),0.1^2\right),
\end{align*}
where $L_{t,1}=1$ if $L_{t-1,1}=1$, and otherwise it is generated from a Bernoulli distribution with parameter $\Psi(-2A_{t-1}+0.2L_{t-1,1})$. For $t=0$, we have $L_{0,1}\sim\calB(0.5)$, and $L_{0,2},L_{0,3}\sim\calN(0,0.1^2)$. The treatment assignment is generated according to the following process:
\begin{align*}
	A_t  \sim \calB\Big(\Psi(-0.5 - L_{t,1}\cos(0.75L_{t,2}) - 0.5L_{t,2}L_{t,3})\Big),
\end{align*}
and the outcome in period $t$ was generated as:
\begin{align*}
	Y_t\sim\calB\Big(\Psi(-2-3A_{t-1}+L_{t-1,1}-6L_{t-1,2}L_{t-1,3}+6L_{t-1,1}L_{t-1,2}^2)\Big),
\end{align*}
for $t=1,\ldots,T$. Two levels of censoring rates were considered:
\begin{align*}
	C_t \sim \calB\Big(\Psi(-\psi_c^*-A_{t-1}+0.75L_{t-1,1}\cos(-0.5L_{t-1,2})-0.5L_{t-1,2}L_{t-1,3})\Big),
\end{align*}
with $\psi_c^*=3$ and 1.5 for 20\% and 50\% censoring, respectively. For evaluation, we consider the treatment strategy $g$ with one tailoring variable
\begin{align*}
	A_t^g = \left\{\begin{array}{l}
		L_{t,2} > 0.2, ~\text{for }t=0\\
		L_{t,2} > 0.2\rvert A^g_{t-1}, ~\text{for }t=1,\ldots,T-1
	\end{array}\right.,
\end{align*}
where one is going to receive the treatment in period $t$ if already treated in period $t-1$ or $L_{t,2}$ exceeds 0.2 in period $t$. For each simulated data, we consider the following methods.
\begin{enumerate}
	\itemsep 0em 
	\item (BART-BS) the proposed BART g-formula estimator with the longitudinal joint score and tailoring variables included, or $b_t=\{e_t,L_{t,2}\}$;
	\item (BART-Cov) the proposed BART g-formula estimator with the actual confounders included in the longitudinal balancing score, or $b_t=L_t$;
	\item (BART-Cov-BS) the proposed BART g-formula estimator with the augmented confounder specification, or $b_t=\{e_t,L_t\}$; 
	\item (NICE) the non-iterative conditional expectation estimator in \citet{Wen2021}, where the BART models in \eqref{eq:bart-oc-mod-simp} and \eqref{eq:bart-long-bs-mod} are replaced by parametric generalized linear models estimated by maximum likelihood;
	\item (IPW) the inverse-probability weighting estimator in \citet{Wen2021};
	\item (AIPW) the multiply-robust augmented inverse probability weighting estimator in \citet{Wen2022sjs};
	\item (TMLE) the targeted maximum likelihood estimator in \citet{Petersen2014}, where the nuisance functions estimated by the \texttt{SuperLearner} with \texttt{R} libraries including \texttt{nnet}, \texttt{nnls}, \texttt{gam}, and \texttt{glmnet} \citep{Lendle2017}.
\end{enumerate}

Additional details, including model specifications of the BART estimators and descriptions of other compared estimators, are given in Web Appendix C. We omit the descriptions of the TMLE estimator for brevity and refer readers to \citet[\S4.2.2]{Petersen2014} for a comprehensive introduction. The NICE, IPW, and AIPW estimators are implemented using \texttt{R} functions available from \citet{Wen2021} and \citet{Wen2022sjs}, and the TMLE is implemented via the \texttt{R} package \texttt{ltmle} \citep{Lendle2017} using their default specification. As detailed in Web Appendix C, the parametric models (to estimate nuisance functions) underlying the NICE, IPW, and AIPW estimators incorporate only linear covariate adjustments. Consequently, they do not fully capture the nonlinear terms and interactions present in the data-generating model. In contrast, TMLE is a frequentist machine learning estimator that uses flexible data-adaptive methods to represent the nuisance functions and is expected to improve upon parametric implementation of NICE, IPW, and AIPW. This design choice serves to illustrate the comparative performance of common estimators under parametric model misspecification. Similar to TMLE, BART also uses flexible models to represent nuisance functions. To implement BART, we draw $15,000$ samples from each posterior distribution, and the first $10,000$ was discarded as burn-in. For each posterior sample of the model parameters, we further generated $10,000$ Monte Carlo samples for the Monte Carlo integration to obtain samples of potential outcomes from the PPD. We replicate the results over 100 synthetic datasets, and values of BART hyper-parameters are set as default, described in Section \ref{subsec:bart-spec}. Simulation results in terms of the relative bias (rBIAS) and root mean squared error (RMSE) across five study periods are summarized in Table \ref{tab:sim}. To aid in the interpretation of the metric, Web Table 1 in Web Appendix C.7 reports the mean squared error (MSE) along with the Monte Carlo standard error for each MSE estimator, as defined in \citet[Table 6]{Morris2019}.
\begin{table*}[htbp]
	\begin{center}
		\caption{Relative bias and RMSE across $T=5$ study periods from the proposed BART estimators and alternative estimators, based on 100 synthetic data replicates.} \label{tab:sim}
		\vspace{0.4cm}
		\resizebox{\linewidth}{!}{
		\begin{tabular}{l rc rc rc rc rc}
			\hline
			& \multicolumn{10}{c}{Censoring rate $=20\%$}\\
			& \multicolumn{2}{c}{$t=1$} & \multicolumn{2}{c}{$t=2$} & \multicolumn{2}{c}{$t=3$} & \multicolumn{2}{c}{$t=4$} & \multicolumn{2}{c}{$t=5$}\\
			\cline{2-11}
			& \multicolumn{1}{c}{rBIAS} & RMSE & \multicolumn{1}{c}{rBIAS} & RMSE & \multicolumn{1}{c}{rBIAS} & RMSE & \multicolumn{1}{c}{rBIAS} & RMSE & \multicolumn{1}{c}{rBIAS} & RMSE \\
			\cline{2-11}
			BART-BS & $-$.026 & .024 & $-$.040 & .035 & $-$.030 & .038 & $-$.023 & .042 & $-$.122 & .091 \\
			BART-Cov & $-$.033 & .037 & $-$.052 & .041 & $-$.023 & .035 & .137 & .085 & .082 & .080 \\
			BART-Cov-BS & $-$.037 & .034 & $-$.053 & .040 & $-$.014 & .031 & .141 & .078 & .087 & .076 \\
			NICE & $-$.075 & .018 & $-$.084 & .031 & $-$.047 & .025 & .613 & .303 & .519 & .321 \\
			IPW & .004 & .013 & $-$.371 & .125 & .020 & .066 & .210 & .140 & .351 & .239 \\
			AIPW & $-$.006 & .016 & .028 & .026 & $-$.201 & .086 & $-$.163 & .087 & $-$.159 & .143 \\
			TMLE & .007 & .012 & $-$.075 & .030 & .022 & .046 & .168 & .117 & .159 & .152 \\
			\hline
			& \multicolumn{10}{c}{Censoring rate $=50\%$}\\
			& \multicolumn{2}{c}{$t=1$} & \multicolumn{2}{c}{$t=2$} & \multicolumn{2}{c}{$t=3$} & \multicolumn{2}{c}{$t=4$} & \multicolumn{2}{c}{$t=5$}\\
			\cline{2-11}
			& \multicolumn{1}{c}{rBIAS} & RMSE & \multicolumn{1}{c}{rBIAS} & RMSE & \multicolumn{1}{c}{rBIAS} & RMSE & \multicolumn{1}{c}{rBIAS} & RMSE & \multicolumn{1}{c}{rBIAS} & RMSE \\
			\cline{2-11}
			BART-BS & $-$.040 & .029 & $-$.035 & .036 & $-$.017 & .040 & $-$.030 & .044 & $-$.125 & .089 \\
			BART-Cov & $-$.029 & .042 & $-$.055 & .043 & $-$.031 & .039 & .144 & .088 & .089 & .082 \\
			BART-Cov-BS & $-$.036 & .035 & $-$.047 & .045 & .002 & .029 & .123 & .086 & .084 & .083 \\
			NICE & $-$.053 & .016 & $-$.104 & .039 & $-$.079 & .039 & .557 & .276 & .499 & .309 \\
			IPW & .007 & .017 & $-$.361 & .121 & .072 & .075 & .398 & .227 & .422 & .283 \\
			AIPW & .008 & .015 & .041 & .027 & $-$.188 & .082 & $-$.146 & .088 & $-$.141 & .154 \\
			TMLE & .009 & .017 & $-$.062 & .029 & .001 & .050 & .093 & .114 & .097 & .183 \\
			\hline
		\end{tabular}
	}
	\end{center}
\end{table*}

Several observations emerge from Table \ref{tab:sim}. First, the NICE estimator frequently has the largest bias, suggesting that the usual implementation of this estimator is likely to suffer from model misspecification. Second, the IPW, AIPW, and TMLE estimators exhibit similar patterns, as they show minimum biases in some or all of the first three periods but could be considerably biased with inflated RMSEs in the last two periods. Among these three estimators, the IPW gives the largest bias and is the least efficient, but it can be improved, as expected, by the AIPW and TMLE with additional information from outcome modeling. Third, across both levels of censoring, although the proposed BART estimators have slightly larger biases and RMSEs in the first two periods, they have mostly smaller biases and are more efficient in later periods. This is likely because flexible nonparametric generative modeling of balancing scores and outcomes becomes increasingly advantageous as cumulative information on outcomes and covariates becomes available in later periods. Finally, we find that the three BART estimators with different choices of balancing scores generally perform similarly with only slight differences. That is, BART-BS tends to perform better in earlier periods due to its parsimonious modeling structure. In contrast, BART-Cov and BART-Cov-BS exhibit smaller biases and lower RMSE in the final period as the availability of streaming information increases. Moreover, BART-Cov-BS appears to enjoy a slight bias and efficiency advantage over the BART-Cov by conducting outcome regression within each longitudinal joint score strata. This observation resembles the finding in a previous simulation with a time-fixed treatment \citep{Hu2021}. However, as shown in Web Table 1, the differences in MSE across different BART implementations sometimes fall within the uncertainty of the simulation experiment.

\subsection{An empirical study with the YNHHS electronic health record data} \label{sec:app}

We demonstrate the implementation of our methods using the Yale New Haven Health System (YNHHS) Electronic Health Records data. The YNHHS data were collected from adult patients admitted to one of the five Yale New Haven health system network hospitals between January 1, 2016, and March 31, 2020, with a length of stay between two to 30 days. Patients hospitalized with a hypertensive emergency or to the maternity ward, intensive care unit, or research unit were excluded \citep{Ghazi2022}. We are interested in studying the effects of antihypertensive treatments on lowering the mean arterial pressure (MAP) for severe hypertension patients within six hours after severe blood pressure elevation (systolic blood pressure [SBP] $> 180$ or diastolic blood pressure [DBP] $> 110$ mm Hg) was reported after admission to the floor. Here, the MAP is defined as $1/3~\mathrm{SBP}+2/3~\mathrm{DBP}$.

The dataset consists of $20,377$ patients, with their baseline characteristics measured at admission: age, biological sex, race, body mass index, admitted ward, admitted hospital, SBP at admission, and DBP at admission. The SBP and DBP of each patient were also monitored every two hours as the study progressed and are considered time-varying covariates. The study duration was discretized into four periods: immediately after severe BP elevation (0), within the first two hours of severe BP elevation ($(0,2)$), between two to four hours ($[2,4)$), and between four to six hours ($[4,6)$). We consider the event of interest as whether the MAP drop $\geq30\%$ compared to that when the severe BP elevation developed, and we are interested in the period to the first event and censor patients whenever the event occurred or the maximum follow-up period is reached. In the dataset, there are $14,197$ (69.7\%) and $1,089$ (5.3\%) patients who were censored and have had the MAP drop $\geq30\%$, respectively. The remaining $5,091$ patients (25.0\%) did not experience the primary event of interest until the end of the follow-up.

For illustration, we compared the effects of different treatment strategies of administrating antihypertensives to severe hypertension patients on lowering their MAP after developing severe BP elevation. We first focused on static deterministic treatment strategies, where the patient was given antihypertensives starting from an initiation period and would continue receiving antihypertensives until the study terminated (a monotone treatment strategy). Four treatment strategies were considered, where the initiation period of antihypertensives ranges from the first to the fourth (last) period; that is, we consider $g\in\mathcal G=\{(1,1,1,1), (0,1,1,1), (0,0,1,1), (0,0,0,1)\}$. Since the outcome event is considered beneficial to patients, the treatment strategy $g$ with the highest $\bbE(Y_4^g)$ is regarded as the most preferable.

\begin{figure}[h]
	\centering
	\includegraphics[width=0.9\textwidth]{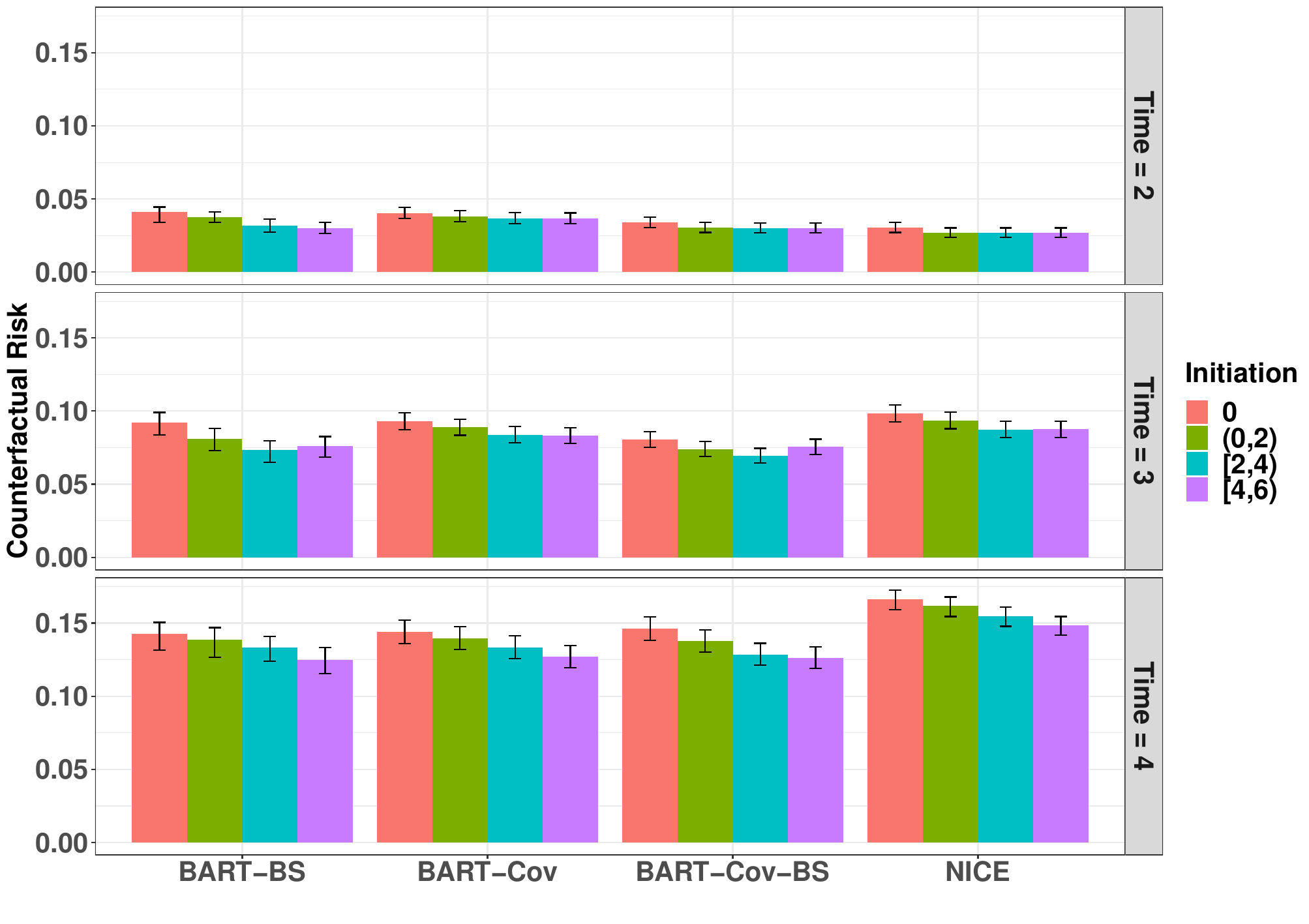}
	\caption{Estimated counterfactual probabilities and 95\% credible intervals of the MAP drop $\geq30\%$ at the last three periods (excluding immediate after the BP elevation) under different treatment strategies defined by the corresponding initiation period of antihypertensives.}
	\label{fig:YNHHS}
\end{figure}

The proposed BART g-formula estimators are used in the analysis of the observational data, and we include the following baseline confounders: age, biological sex, race, the body mass index, admitted ward, admitted hospital, SBP at admission, and DBP at admission, as well as two time-varying confounders: SBP and DBP. To implement the BART estimators (BART-BS, BART-Cov, BART-Cov-BS), we draw $25,000$ samples from each posterior distribution with the first $15,000$ discarded as burn-in, and for each posterior sample of the model parameters, we generated $10,000$ samples for the Monte Carlo integration to obtain samples of potential outcomes from its PPD. The model specifications align closely with those in Web Appendices C.1 - C.3. For a reference comparison, we also applied the parametric NICE estimator. Specifically, let $\tilde L_0$ denote the vector of all baseline characteristics, and then separate linear regression models were fitted for $\mathrm{SBP}_t$ with $(\mathrm{SBP}_{t-1}, A_{t-1}, \tilde L_0')'$ as predictors, and $\mathrm{DBP}_t$ with $(A_{t-1},\mathrm{DBP}_{t-1},\mathrm{SBP}_t,\tilde L_0')'$ as predictors. A logistic regression model was fitted for the binary outcome $Y_t$ with $(\mathrm{SBP}_t,\mathrm{DBP}_t, A_{t-1},\tilde L_0')'$ as predictors (also see Web Appendix C.4). Results from the BART estimators and the NICE estimator are presented in Figure \ref{fig:YNHHS}.

Across all estimators examined, the results suggest that earlier initiation of antihypertensives could increase the probability of patients experiencing a drop in MAP of 30\% or more within six hours after severe BP elevation occurs. The most preferred treatment strategy is administrating antihypertensives immediately following the onset of severe BP elevation, which is consistently suggested by all methods considered. Comparing results from the NICE estimator with those from the BART estimators, we observe that the NICE estimator gives slightly higher values of $\bbE(Y_4^g)$. However, the results obtained using different specifications of $b_t$ for the BART g-formula are relatively consistent in this single data analysis example. From the patterns observed in our main simulations with synthetic data, this difference might be because the generalized linear models employed in the NICE estimator are misspecified, leading to an overestimation at a later period. As a further illustration, we provide additional analysis results comparing random treatment strategies also using the YNHHS dataset in Web Appendix D.

\section{Discussion} \label{sec:dis}

In this article, we describe a general class of g-formulas for analyzing discrete-time observational survival data with binary time-varying treatments. Our survival g-formula introduces the longitudinal balancing score and allows for a broader range of specifications, serving as a useful alternative to the classic survival parametric g-formula. We show that the coarsest or minimum sufficient longitudinal balancing score, without any tailoring variables, is the joint propensity and censoring score, which is a low-dimensional summary in each period for all time-varying confounders. This result offers modeling convenience, as it allows for modeling a single scalar value rather than a potentially extensive list of time-varying confounders. Furthermore, we propose flexible BART specifications for each component model in the survival g-formula to improve upon the operating characteristics of the standard parametric g-formula.

In our study, the inclusion of the longitudinal joint score primarily arises from reducing the modeling complexity by avoiding the need to model all time-varying covariates, which may be multi-dimensional and is mainly to improve the feasibility of implementing the conventional g-formula estimator \citep{Achy-Brou2010}. From this perspective, our study continues to highlight the use of propensity scores in estimating causal effects. As pointed out in \citet{Li2022}, there has been controversy regarding whether propensity scores should be included in an outcome model for Bayesian causal inference, as the propensity score factors out the joint likelihood under prior independence. However, several previous studies demonstrated the additional benefits of including an estimated propensity score for outcome modeling; see, for example, \citet{little2004robust,zigler2013model,Zhou2019} and \citet{Hu2021}. With a time-fixed treatment, \citet{Hahn2020} developed the Bayesian Causal Forest for estimating the conditional average treatment effect with a point treatment in the cross-sectional setting, which specifies separate BART priors for the baseline mean function (which includes an estimated propensity score) and the conditional average treatment effect function, to address regularization-induced confounding. They have also shown that adding the estimated propensity score in the baseline mean function can improve the empirical estimation of the conditional average treatment effect. 

Based on our Bayesian implementation of the survival g-formula, there are several important directions for future research. First, we have primarily focused on a binary time-varying treatment, whereas the original development of the g-formula can also accommodate a continuous time-varying treatment under slightly modified identification assumptions. Although it is possible to expand Theorems \ref{thm:thm-bs} - \ref{thm:thm-new-gform} to a continuous treatment $A_t$ and generalize \eqref{eq:new-gform} accordingly, the longitudinal joint score may now depend on the conditional density of the continuous treatment in each period and would necessitate additional computational considerations. The finite-sample performance and practice considerations of the new g-formula with a continuous time-varying treatment are an important direction for future research. Second, as in most longitudinal observational studies, we have been operating under the key assumption of sequential ignorability for identifying the causal effects of treatment strategies. Although the sequential ignorability assumption is not verifiable from the observed data alone, an important direction for future investigation is to develop systematic sensitivity strategies to assess the impact under unmeasured baseline and time-varying confounding. With a point treatment, \citet{hu2022flexible} recently developed a sensitivity function approach to bias-correct the observed outcome under unmeasured baseline confounding and considered BART for coherent posterior inference under violation of the unconfoundedness assumption. With a time-varying treatment, sensitivity methods that address unmeasured confounding and outcome misclassification bias for parametric g-formulas have been studied by \citet{danaei2016weight,bijlsma2019impact} and \citet{Edwards2018}. It would be of substantial interest and value to develop these existing sensitivity methods further under the BART implementation of the proposed survival g-formula.

Third, while allowing for flexible estimation of nuisance parameters, when grounded in a frequentist statistical model, the proposed BART estimators do not solve an efficient influence function or share the corresponding properties of double machine learning (DML) or TMLE methods \citep{Petersen2014,Chernozhukov2018}. With a time-fixed treatment, \citet{Naimi2021} have shown via simulations that singly robust g-formula estimators with machine learning methods (random forests, extreme gradient boosting, generalized additive models) may not improve the estimation of causal effects over misspecified parametric models, and both are improved by DML or TMLE. In our more complex time-varying treatment scenario, we find that BART estimators can still improve upon misspecified parametric estimators, although they only become less biased and more efficient than AIPW and TMLE at later periods as the availability of streaming information increases. This, along with prior simulations that recommended BART over other machine learning methods \citep{Dorie2019,hu2020estimation,caron2022estimating}, may suggest that BART remains a competitive data-adaptive approach for causal inference. However, as pointed out by \citet{hill2020bayesian}, there are open questions to solve regarding the asymptotic theory of BART, and hence the frequentist inferential properties of BART causal estimators require further investigation. A promising future direction is to incorporate BART as a data-adaptive method to learn the nuisance functions of the efficient influence function tailored to the target causal estimand. Assuming a point treatment, \citet{antonelli2022causal} developed an estimator that combines the strengths of frequentist and Bayesian methods for causal inference and proposed a valid confidence interval estimator. Extending their estimator to accommodate time-varying treatments could lead to an estimator that combines the advantages of BART with rate-doubly or multiply robust methods, offering improvements over the current BART-based g-formula estimators.

\bibliography{gform}

\end{document}